# Non-Abelian charged nodal links in dielectric photonic crystal

Haedong Park[1], Stephan Wong[1], Xiao Zhang[2], Sang Soon Oh[1,*]

[1]School of Physics and Astronomy, Cardiff University, Cardiff CF24 3AA, United Kingdom

[2]School of Physics, Sun Yat-sen University, Guangzhou 510275, China

*Email: ohs2@cardiff.ac.uk

**Abstract**: A nodal link is a special form of a line degeneracy (a nodal line) between adjacent bands in the momentum space of a three-dimensional topological crystal. Unlike nodal chains or knots, a nodal link consists of two or more mutually-linked rings that do not touch each other. Recent studies on non-Abelian band topology revealed that the topological charges of the nodal links can have the properties of quaternions. However, a photonic crystal that has a nodal link with non-Abelian charges has not been reported. Here, we propose dielectric photonic crystals in forms of double diamond structures which realize the nodal links in the momentum space. By examining the evolution of the eigenstate correlations along the closed loops which enclose the nodal line(s) of the links, their non-Abelian topological charges are also analyzed. The proposed design scheme and theoretical approach in this work will allow for experimental observation of photonic non-Abelian charges in purely dielectric materials and facilitate the control of the degeneracy in complex photonic structures.

**Main text**:

Topological crystals generally deal with degenerate states in the band structures. In many cases, the dimensions of the degeneracies are zero-dimensional (0D) such as Dirac[1-4] or Weyl points[5-13], or one-dimensional (1D) such as nodal lines[14-18]. Such topological crystals can induce interesting boundary states such as one-way edge/surface states[19-21], surface localizations[22-24], arc surface states[25-28], drumhead surface states[29-33], and bound states in the continuum (BIC)[34]. The nodal lines, protected by the product of time-reversal symmetry (T) and parity (P) inversion[35], are categorized as nodal rings[29,30], nodal chains[31,32,36,37], nodal links[31-33,37-41], and nodal knots[33,41,42]. A nodal ring is a simple closed-loop form of the nodal line. Two or more nodal rings form a nodal chain when any two rings of them are connected at a single point[31,32,36,37]. If two or more nodal rings are connected without any touch, they form a nodal link, e.g., the Hopf link[31-33,37-41]. A nodal knot is a single closed loop which cannot be deformed into the unknot state by Reidemeister moves, e.g., the trefoil knot[33,41,42]. Extensive efforts have been recently made to find nodal lines in topological metals[15,40], semimetals[14,31,37,39,41,42], phononic crystals[29], and electrical circuits[33]. A number of studies have discussed topological invariants of the nodal lines such as the non-Abelian Berry-Wilczek-Zee (BWZ) connection[43], winding number[41,44], and Chern number[44]. Very recently, a study on non-Abelian band topology showed that the topological charges of the nodal lines including the nodal links or nodal chains can be described as quaternions[15].

Despite the abundance of studies on the nodal lines[13,16,30,36,38] and the non-Abelian band topology in photonics[38], there has not been any textbook example on the nodal link. A photonic crystal can be such an example if (i) its output nodal lines completely satisfy the aforementioned definition of the nodal link[41] and (ii) the nodal links of the photonic crystal have non-Abelian topological charges[15]. The existing photonic crystals on the nodal lines do not completely fulfil at least one of these. In particular, although the second condition has been dealt with using the effective 3×3 Hamiltonian derived from the Maxwell's equations[38], there is not any precedent in which the non-Abelian topological charges are calculated based on the full-vector field form of eigenstates computed for a photonic crystal. Additionally, most of the existing photonic studies on the nodal line[16], nodal ring[30], nodal chain[36], or nodal links[38] are commonly limited to the metallic structures with mirror symmetry.

In this work, we theoretically demonstrate the nodal link with non-Abelian charges by using a dielectric photonic crystal. First, we introduce our double diamond structure which has inversion symmetries but does not have mirror symmetry. Then, the nodal link in the three-dimensional (3D) momentum space of this structure is discussed. The non-Abelian topological features of the nodal link are characterized by considering a loop that encloses one or more sections of the link and examining the correlations of the eigenstates.

The realization of the nodal link in a dielectric photonic crystal starts with taking the well-known diamond structure[45] and subsequently breaking its several geometrical symmetries as will be explained below. For the face-centered cubic (FCC) primitive cell whose lattice vectors are $\mathbf{a}_1 = a/2\,[0, 1, 1]$, $\mathbf{a}_2 = a/2\,[1, 0, 1]$, and $\mathbf{a}_3 = a/2\,[1, 1, 0]$, our bicontinuous photonic crystal is defined by a set of $\mathbf{x} = [x_1, x_2, x_3]$ such that $f(\mathbf{x}) > f_c > 0$ or $f(-\mathbf{x}) > f_c > 0$ where $f(\mathbf{x})$, the triply periodic level surface, is expressed as:

$$f(\mathbf{x}) = \sin(X_1 + X_2 + X_3) + \sum_{i=1}^{3} A_i \sin(X_1 + X_2 + X_3 - 2X_i) \qquad (1)$$

Here, $\mathbf{X} = [X_1, X_2, X_3] = (2\pi/a)(\mathbf{x} - \gamma \mathbf{a}/2)$ is a normalized local coordinate defined with the lattice constant $a$, global coordinates $\mathbf{x}$, the summation of the lattice vectors $\mathbf{a} = \sum_{i=1}^{3} \mathbf{a}_i$, and the coefficient $\gamma$ which describes the translation of one diamond from the other along the [1, 1, 1]-direction. By the inequalities mentioned above, this photonic crystal consists of two inversion symmetric single diamonds and they do not intersect each other, like other bicontinuous structures[5,12,13,46-49]. If $A_1 = A_2 = A_3 = 1$, the space group of each part becomes $Fd\bar{3}m$ (No. 227) and each diamond is identical to the well-known diamond structure[45]. If $\gamma = 0$ is also satisfied, the space group of the total structure becomes $Pn\bar{3}m$ (No. 224) whose lattice constant is only $a/2$[46-49]. Under this condition, a perfect nodal link is not observed because of extra degeneracies (see Supplementary Section S1). To remove the redundant degeneracies, several geometrical symmetries should be broken by the following two modifications. Setting the coefficients $A_i$ different to each other and not equal to 1 destroys all the symmetries except translational and inversion symmetries and makes the structure anisotropic along all directions. Under this symmetry breaking, the two single diamonds are still mutual-inversion symmetric and each diamond is also self-inversion symmetric that the inverse of a single diamond coincides itself. Here, the self-inversion symmetric points of the two single diamonds are identical. The second modification is setting nonzero $\gamma$ which splits the self-inversion symmetric points along [1, 1, 1] -direction. This also eliminates the translational symmetry with the unit vector $\langle a/2, 0,0 \rangle$ so that not the primitive cubic cell with $(a/2)^3$ but the above FCC cell has to be used (the detailed analyses on the symmetries of the double diamond structures are discussed in Supplementary Section 2). The mutual-inversion symmetry by two single diamond structures survives which is a necessary condition for the formation of a nodal link[13,35]. Each diamond structure in these configurations is depicted in Fig. 1a; the sets of **x** which satisfies $f(\mathbf{x}) > f_c$ and $f(-\mathbf{x}) > f_c$ are given as the pink and sky-blue colored structures, respectively.

The 3D photonic band structure for the double diamond photonic crystal has multiple degeneracies that form a link. To probe all the degeneracies, we calculated the photonic band structure by using the MIT Photonic-Bands (MPB) package[50]. We regarded any point as degenerated by adjacent two bands if the normalized frequency difference ($\Delta\omega a/2\pi c$) of these bands at the point is smaller than a critical tolerance, 0.0045. Sets of the degeneracies formed by the 3rd and 4th bands and 4th and 5th bands appear as rings, depicted as orange and cyan shapes, respectively, in Fig. 1b. These two rings are mutually linked like a ship chain. Features of this result are as follows: (i) This link is infinitely repeated along the connection direction because of the periodicity of the first Brillouin zone. The cyan ring centered at Γ-point connects the orange rings spanning the zone boundary. (ii) The link is formed not along but avoiding the high symmetry lines. To see this, we first set three paths along the high symmetry directions of the FCC cell; all three paths depart from their own points on the zone boundary and, via the Γ-point, go to other points on the zone boundary. Then, we drew the band structures along these paths as shown in Fig. 1c-e. The $U_a\Gamma$ and $\Gamma U_b$ of $U_a\Gamma U_b$ pass close by the orange and cyan rings, respectively (see Fig. 1c). But the 3rd, 4th, and 5th bands do not completely touch along this path; $U_a\Gamma U_b$ does not exactly meet the link. The results in Fig. 1d-e can be explained in the same way. As such, these three bands commonly show few degeneracy-like points. This contrasts to the case of the photonic crystal with $A_1 = A_2 = A_3 = 1$ and $\gamma = 0$ (see Supplementary Section S1); its resulting band structures show high similarity between 3rd and 4th bands and the degeneracies between 4th and 5th bands always appear around the Γ-point. For extra degeneracies outside the link shown in Fig. 1b, see Supplementary Section S3. In addition to the nodal link in Fig. 1b, the set of degeneracies by the 1st and 2nd bands forms a nodal chain, as shown in Supplementary Section S4.

The topological charges of the link can be described by the non-Abelian quaternion group

$\mathbb{Q} = \{1, \mathbf{i}, \mathbf{j}, \mathbf{k}\}$[15,38]. The last three numbers are defined such that $\mathbf{i}^2 = \mathbf{j}^2 = \mathbf{k}^2 = -1$ and their relations are skew-symmetric; $\mathbf{ij} = -\mathbf{ji} = \mathbf{k}$, $\mathbf{jk} = -\mathbf{kj} = \mathbf{i}$, and $\mathbf{ki} = -\mathbf{ik} = \mathbf{j}$. To observe the topological charges, the 3rd, 4th, and 5th bands are labelled as $n = 1, 2$, and $3$, respectively. A closed loop that encircles a point on the orange ring (by the bands $n = 1$ and $2$) is then considered (Fig. 2a). We define the correlation $\mathbf{C}_n(\mathbf{k})$ of the band $n$, also known as the polarization[38], at any point $\mathbf{k}$ on the closed loop as

$$[\mathbf{C}_n(\mathbf{k})]_m = \langle u_{\mathbf{k}_0}^m | u_{\mathbf{k}}^n \rangle = \int_{cell} \left(u_{\mathbf{k}_0}^m\right)^* \cdot u_{\mathbf{k}}^n d^3\mathbf{x} \qquad (2)$$

where $\mathbf{k}_0$ means the starting point of the closed loop (marked as a circle in Fig. 2a) and $u_{\mathbf{k}}^n$ is a periodic part of the magnetic field eigenstate $\mathbf{H}^n(\mathbf{x}) = u_{\mathbf{k}}^n e^{i\mathbf{k}\cdot\mathbf{x}}$ of the band $n$ such that $\langle u_{\mathbf{k}}^m | u_{\mathbf{k}}^n \rangle = \delta_{mn}$. In other words, the correlation of the band $n$ is calculated from the projection of $|u_{\mathbf{k}}^n\rangle$ onto $|u_{\mathbf{k}_0}^m\rangle$. (A preparation on the sign convention for $|u_{\mathbf{k}}^n\rangle$ is needed to use equation (2) and this is discussed in Supplementary Section S5.) After one winding along the loop, the correlations of the bands $n = 1$ and $2$ exhibit $\pi$-rotations while the correlations of the band $n = 3$ do not change (Fig. 2b). Gathering the tails of all these arrows into the origin generates Fig. 2c; only $\mathbf{C}_3$ remains fixed while $\mathbf{C}_1$ and $\mathbf{C}_2$ rotate by $\pi$ around $\mathbf{C}_3$. For $\alpha \in [0, 2\pi]$ which parametrizes the closed loop, the rotation matrix for Fig. 2c is expressed as $R_{12}(\alpha) = e^{(\alpha/2)L_3}$ where $(L_i)_{jk} = -\epsilon_{ijk}$. Here, the subscript $12$ means that the loop encloses the ring formed by the bands $n = 1$ and $2$. Rewriting $R_{12}(\alpha)$ by its lift in the double cover Spin(3), i.e. $\bar{R}_{12}(\alpha) = e^{-i(\alpha/2)(\sigma_3/2)}$, with now $\alpha \in [0, 4\pi]$ and substituting $\alpha = 2\pi$ gives the topological charge $-i\sigma_3$ where $\sigma_i$ are the Pauli matrices[15]. Similar analysis can be carried out on a closed loop which encircles a cyan ring (by the bands $n = 2$ and $3$) as shown in Fig. 2d. In this case, the correlations of the bands $n = 2$ and $3$ show the $\pi$-rotations and only the correlations of the band $n = 1$ keep the initial direction (Fig. 2e-f). Thus, the topological charge $\bar{R}_{23}(2\pi)$ for Fig. 2f is $-i\sigma_1$.

Meanwhile, if we consider a loop that encloses both orange and cyan rings (Fig. 3a), the $\pi$-rotations occur for the correlations of the bands $n = 1$ and $3$ (Fig. 3b). In contrast to the case of $-i\sigma_3$ and $-i\sigma_1$ (Fig. 2c and f), the correlations in Fig. 3c reveals a more complicated evolution where $\mathbf{C}_1(\mathbf{k})$ and $\mathbf{C}_3(\mathbf{k})$ do not rotate on a single plane ($\hat{\mathbf{x}}_1\hat{\mathbf{x}}_3$-plane) and $\mathbf{C}_2(\mathbf{k})$ is not fixed along a specific direction ($\hat{\mathbf{x}}_2$-axis). This can be simplified by aligning all $\mathbf{C}_2(\mathbf{k})$ along $\hat{\mathbf{x}}_2$-axis as follows. Let us consider $\mathbf{C}_n$ at a point $\mathbf{k}$ on the loop (Fig. 3d). We can assume a rotation matrix $\mathbf{R}_{\mathbf{C}_2 \to \hat{\mathbf{x}}_2}(\mathbf{k}, \theta)$ which rotates $\mathbf{C}_2(\mathbf{k})$ through an angle $\theta$ with respect to the axis $\mathbf{r}(\mathbf{k}) = \mathbf{C}_2(\mathbf{k}) \times \hat{\mathbf{x}}_2$. By applying $\mathbf{R}_{\mathbf{C}_2 \to \hat{\mathbf{x}}_2}(\mathbf{k}, \theta)$ to not only $\mathbf{C}_2(\mathbf{k})$ but also $\mathbf{C}_1(\mathbf{k})$ and $\mathbf{C}_3(\mathbf{k})$, the correlations in Fig. 3c can be calibrated. If the angle between $\mathbf{C}_2(\mathbf{k})$ and $\hat{\mathbf{x}}_2$ is denoted as $\theta_0(\mathbf{k})$, the correlations transformed by $\mathbf{R}_{\mathbf{C}_2 \to \hat{\mathbf{x}}_2}(\mathbf{k}, 0.6\theta_0)$ and $\mathbf{R}_{\mathbf{C}_2 \to \hat{\mathbf{x}}_2}(\mathbf{k}, \theta_0)$ are shown in Fig. 3e and f, respectively. The topological charge $\bar{R}_{31}(2\pi)$ for the loop that encloses the two rings is therefore $-i\sigma_2$. Because the closed loop first circles around the cyan node ($-i\sigma_1$) and then around the orange node ($-i\sigma_3$), the resultant topological charge shown in Fig. 3 satisfies the relation $-i\sigma_2 = (-i\sigma_3)(-i\sigma_1)$. If this sequence is reversed, the sign of this charge may also be flipped. This is discussed in Supplementary Section S6. Meanwhile, the topological charge in Fig. 3 also can be drawn without the above calibration and the discussions are in Supplementary Section S7.

Besides, the topological charges $I$ and $-I$ can be deduced from the above results. Selecting the same kind of nodes (i.e., same-colored nodes) and setting the closed loop which ties oppositely oriented nodes will generate $(-i\sigma_1)(i\sigma_1) = (-i\sigma_3)(i\sigma_3) = I$. If the loop ties nodes with the same orientation, the result will be $(-i\sigma_1)(-i\sigma_1) = (-i\sigma_3)(-i\sigma_3) = -I$. The former case corresponds to $\alpha$ which increases from zero to $2\pi$ followed by decreasing to zero. Applying this path to the orange ring makes $C_1$ and $C_2$ rotate by $\pi$ for $\alpha \in [0, 2\pi]$ and by $-\pi$ for $\alpha \in [2\pi, 0]$ while keeping $C_3$ (Fig. 4a). The net evolutions of both $C_1$ and $C_2$ are zero and therefore the topological charge becomes $I$. The latter case is expressed by setting $\alpha \in [0, 4\pi]$. Along the path, $C_1$ and $C_2$ rotate by $2\pi$ around $C_3$ as shown in Fig. 4b. This situation is simply written by $\bar{R}_{12}(\alpha)$, as mentioned regarding Fig. 2c in the previous paragraph; substituting $\alpha = 4\pi$ into $\bar{R}_{12}(\alpha)$ gives the topological charge $-I$. Same analyses can be carried out on the cyan ring if only $C_1$, $C_2$, and $C_3$ in the above sentences are replaced as $C_2$, $C_3$, and $C_1$, respectively, to generate the results shown in Fig. 4c-d.

All these topological charges can be replaced as quaternions by $-i\sigma_3 \mapsto \mathbf{k}$, $-i\sigma_1 \mapsto \mathbf{i}$, $-i\sigma_2 \mapsto \mathbf{j}$, $I \mapsto 1$, and $-I \mapsto -1$ because the nature of $\{I, -i\sigma_1, -i\sigma_2, -i\sigma_3\}$ is isomorphic to $\mathbb{Q}$. Therefore, a closed loop which encloses the ring(s) of the nodal link is classified under the quaternion group. (Another method of determining the topological nature of the rings is using the non-Abelian BWZ connection[43] and this is discussed in Supplementary Section S8.)

The existence of the nodal link may open the possibility of observing the topological surface states in our double diamond structure[30,31,36]. To explore this, the unit cells shown in Fig. 1a are stacked along $\mathbf{a}_2$-direction. The perfect electric conductor (PEC) is imposed on both boundaries at the ends of this stacked array. The boundaries are parallel to $\mathbf{a}_1$ and $\mathbf{a}_3$-directions and these $\mathbf{a}_1$ and $\mathbf{a}_3$ are the same directions as $\Gamma K_1$ and $\Gamma K_3$-directions marked in Fig. 5a, respectively. The band structure along the $\Gamma \bar{K}_m$-direction shows the bands (the two red-colored and overlapped bands in Fig. 5b) between bulk states. These bands are connected to $\bar{A}_L$ at $\omega_L$ which are the frequency and point at which the cyan nodal line passes, respectively. The localizations of $\|\mathbf{H}(\mathbf{x})\|$ around the boundaries are observed as shown in Fig. 5c and these demonstrate that the isolated bands in Fig. 5b are therefore the topological surface states.

In summary, we have demonstrated the non-Abelian topological nodal link in the momentum space using the double diamond photonic crystal made of dielectric material. The double diamond structure was prepared by breaking geometrical symmetries of the double diamond structure in the space group of $Pn\bar{3}m$ (No. 224) except translational and inversion symmetries. We have shown that the reduction of symmetries removes redundant degeneracies that exist in the original double diamond structure. With the correlation defined from the full vector-field eigenstates and the closed-loop analyses, we have shown that their topological charges are non-Abelian and belongs to the quaternion group. Importantly, our results clearly show that the realization of non-Abelian topological nodal link is not limited by the symmetry and material configurations used in the previous studies on the nodal nodes including nodal links[31-33,37-41], nodal chains[31,32,36,37] and nodal knots[33,41,42]. We believe that our results will lay out the theoretical ground for the experimental observation of photonic non-Abelian charges in purely dielectric materials and allow more freedom in controlling the degeneracy in complex photonic structures.

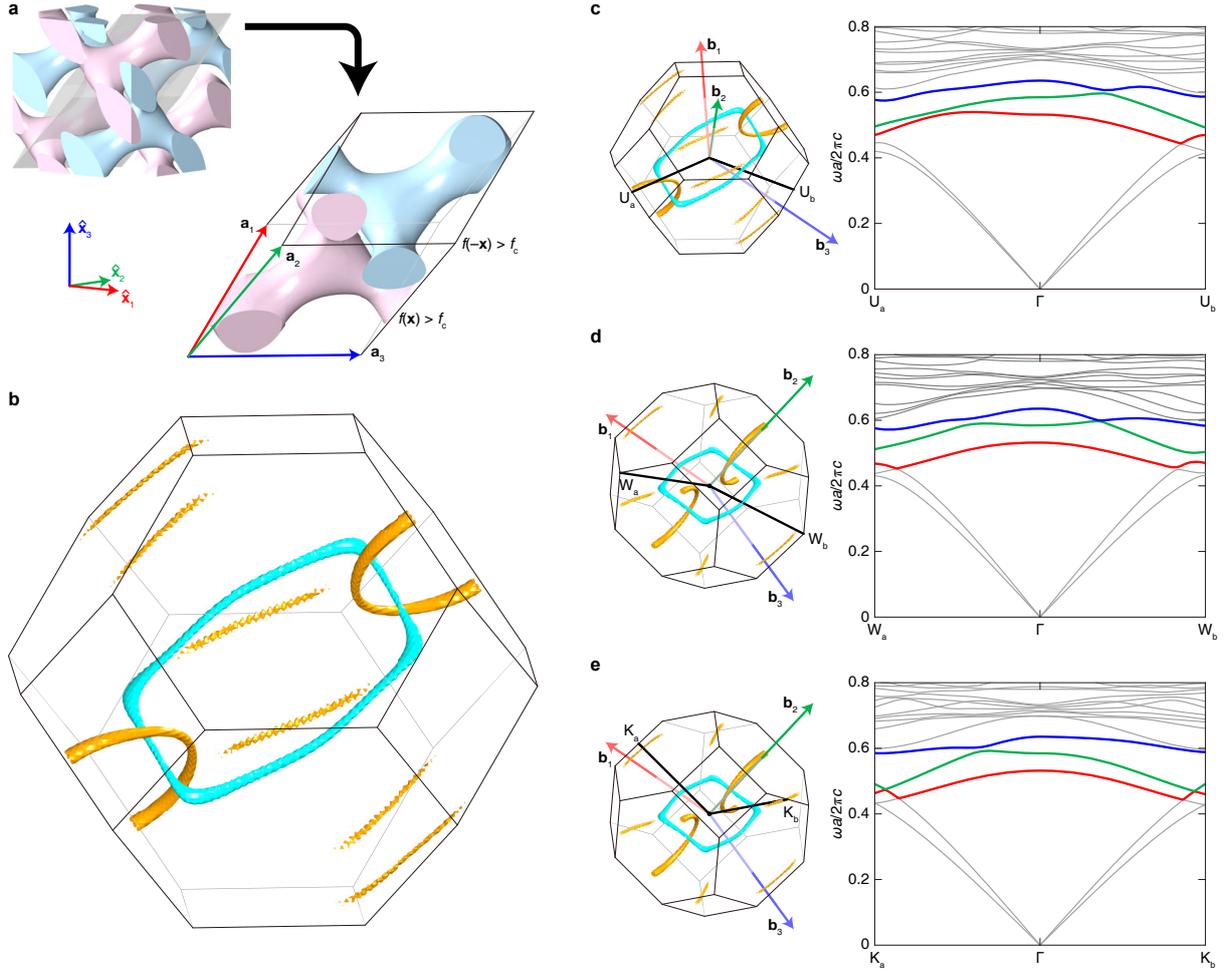

**Fig. 1. Double diamond photonic crystal and its nodal link. a**, Real space geometry with **A** = [1.19, 1.37, 1.28], $\gamma = 0.08$, and $f_c = 1.85$. Dielectric permittivity is 16.0 for both structures. The FCC primitive cell is defined by $\mathbf{a}_1 = a/2\,[0,1,1]$, $\mathbf{a}_2 = a/2\,[1,0,1]$, and $\mathbf{a}_3 = a/2\,[1,1,0]$. **b**, Nodal link in 3D momentum space. The orange and cyan nodal rings are respectively formed by the 3rd and 4th bands and 4th and 5th bands in the band structure. **c-e**, Band structures (right) along the several paths (left).

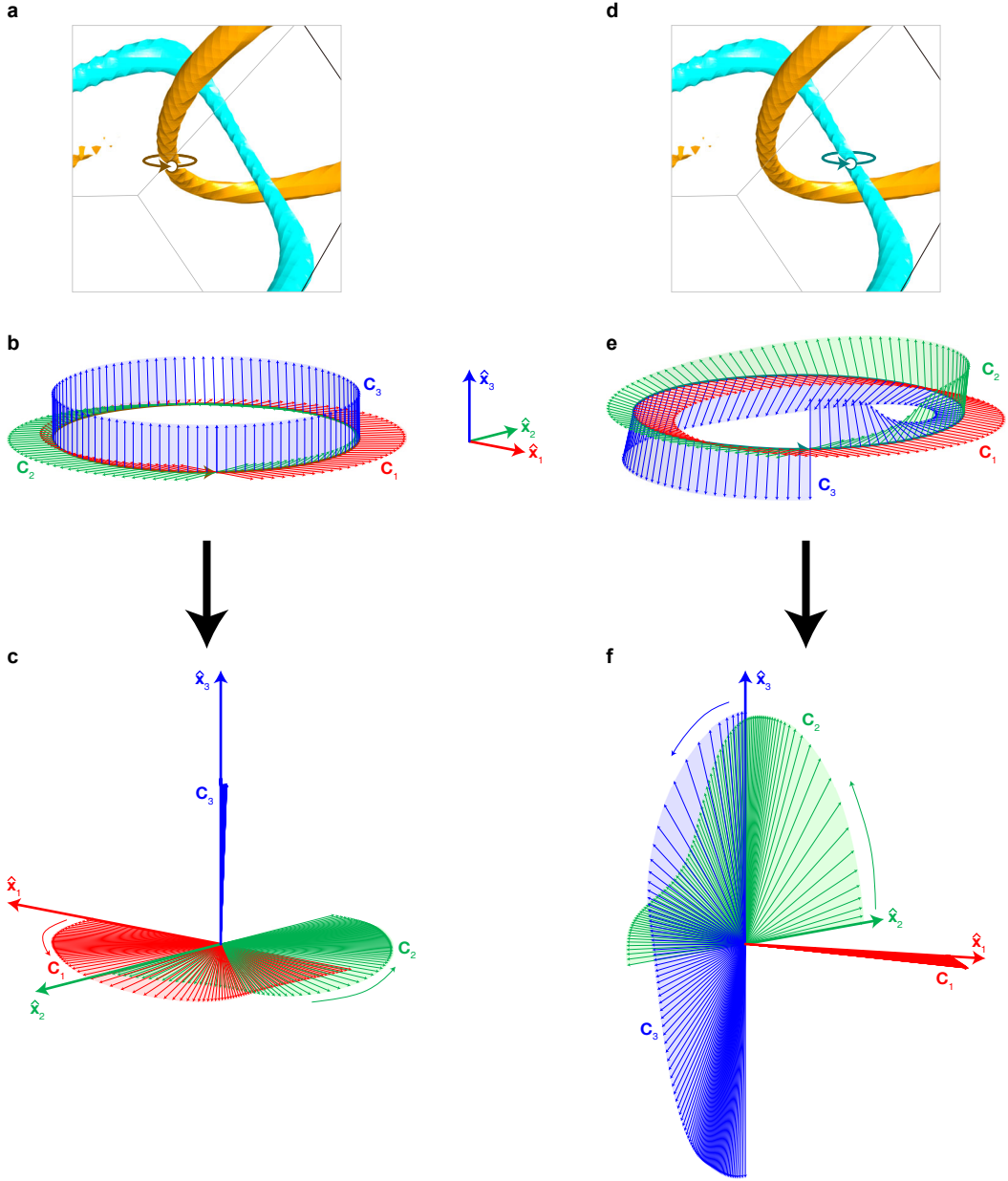

**Fig. 2. Derivation of the topological charges i and k. a** and **d**, Closed loops which enclose the orange and cyan nodal rings, respectively. Their winding directions and starting points $\mathbf{k}_0$ are also marked as arrows and circles, respectively. **b** and **e**, Correlations $\mathbf{C}_n$ from the eigenstates, along the closed loops. Only their real parts are plotted. **c** and **f**, Correlations $\mathbf{C}_n$ whose vector tails are collected at the origin to see their topological charges. In **b-c** and **e-f**, we set the arbitrary orthonormal directions as $\hat{\mathbf{x}}_1$, $\hat{\mathbf{x}}_2$, and $\hat{\mathbf{x}}_3$.

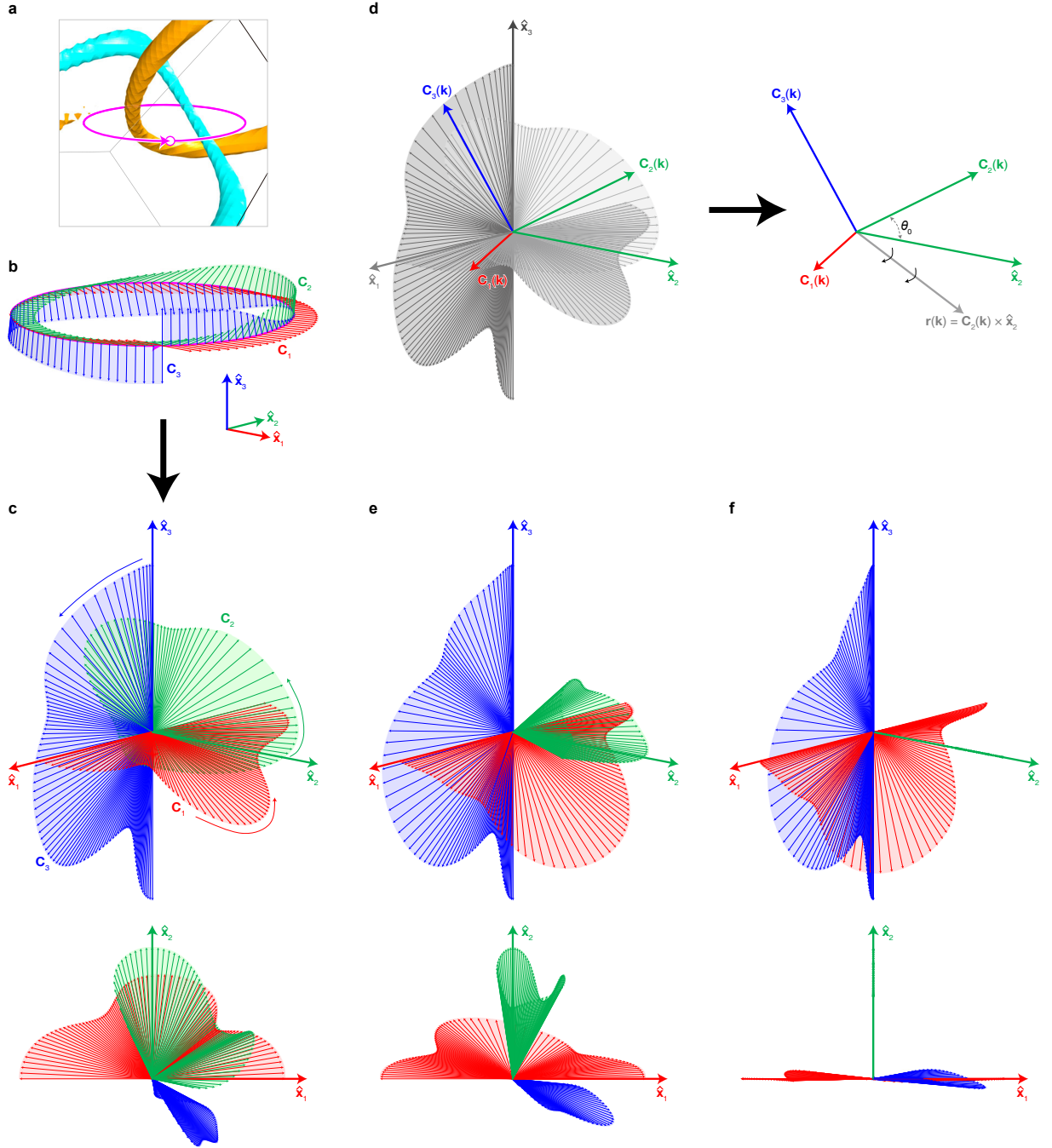

**Fig. 3. Derivation of the topological charge j. a**, Closed loop which encloses the cyan and orange nodal rings. Its winding direction and starting point are also marked as an arrow and a circle, respectively. **b** and **c**, Correlations $\mathbf{C}_n$ plotted by same method in Fig. 2. **d**, $\mathbf{C}_n(\mathbf{k})$ at an arbitrary $\mathbf{k}$ and $\hat{\mathbf{x}}_2$-axis. The matrix $\mathbf{R}_{\mathbf{C}_2 \to \hat{\mathbf{x}}_2}(\mathbf{k}, \theta)$ is defined such that this matches $\mathbf{C}_n(\mathbf{k})$ onto $\hat{\mathbf{x}}_2$-axis with respect to $\mathbf{r}(\mathbf{k})$ perpendicular to both $\mathbf{C}_n(\mathbf{k})$ and $\hat{\mathbf{x}}_2$. **e** and **f**, Calibration results of **c** by $\mathbf{R}_{\mathbf{C}_2 \to \hat{\mathbf{x}}_2}(\mathbf{k}, 0.6\theta_0)$ and $\mathbf{R}_{\mathbf{C}_2 \to \hat{\mathbf{x}}_2}(\mathbf{k}, \theta_0)$, respectively, where $\theta_0$ is the angle between $\mathbf{C}_n(\mathbf{k})$ and $\hat{\mathbf{x}}_2$.

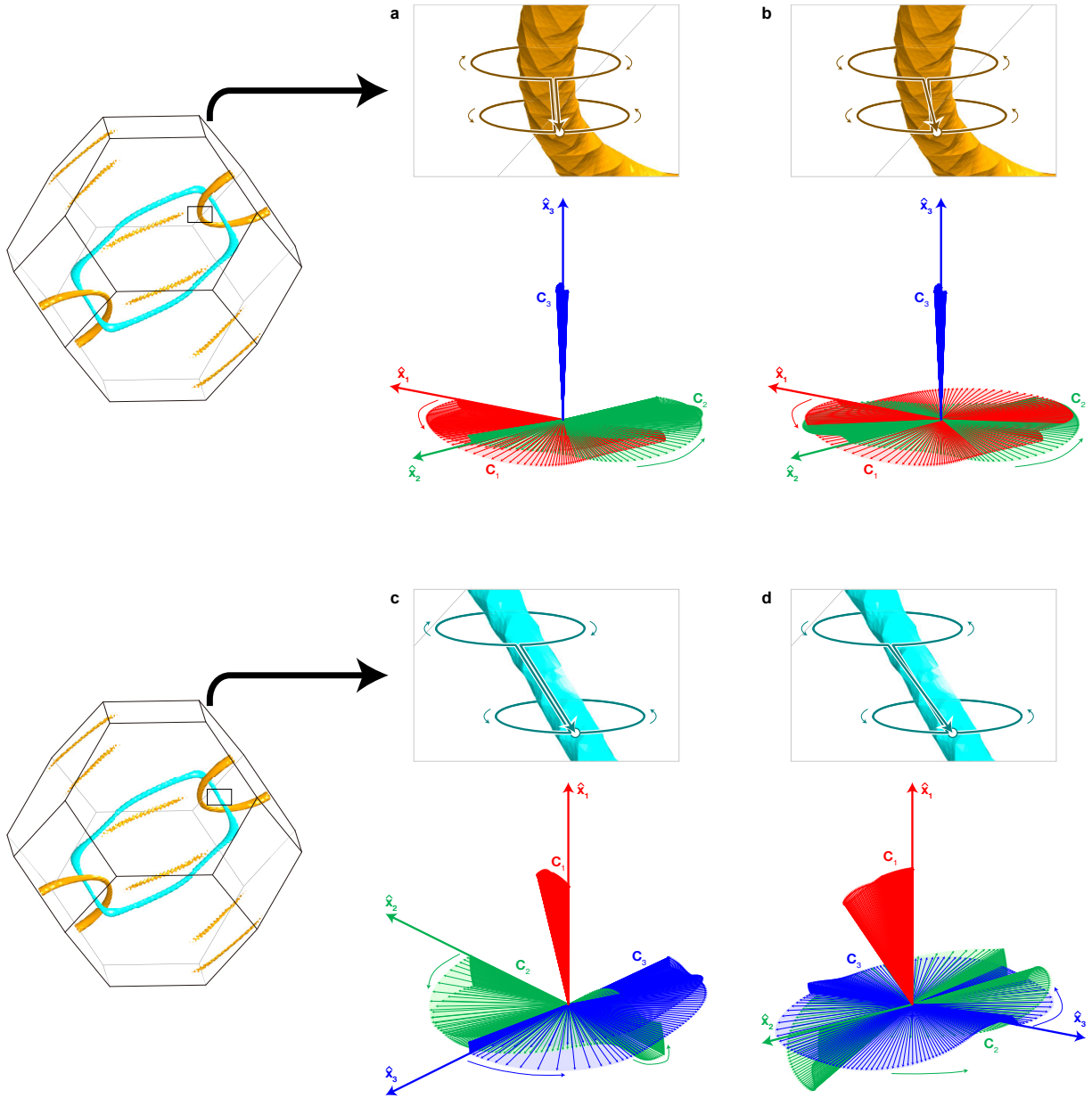

**Fig. 4. ±1 Topological charges. a-d**, Correlations $C_n$ of the eigenstates along the loops which enclose the orange (**a-b**) and cyan (**c-d**) nodes. Only the real parts of the correlations are plotted. All these four panels have two circles in the enlargement insets. The winding directions of these two circles are opposite in **a** and **c** and same in **b** and **d**.

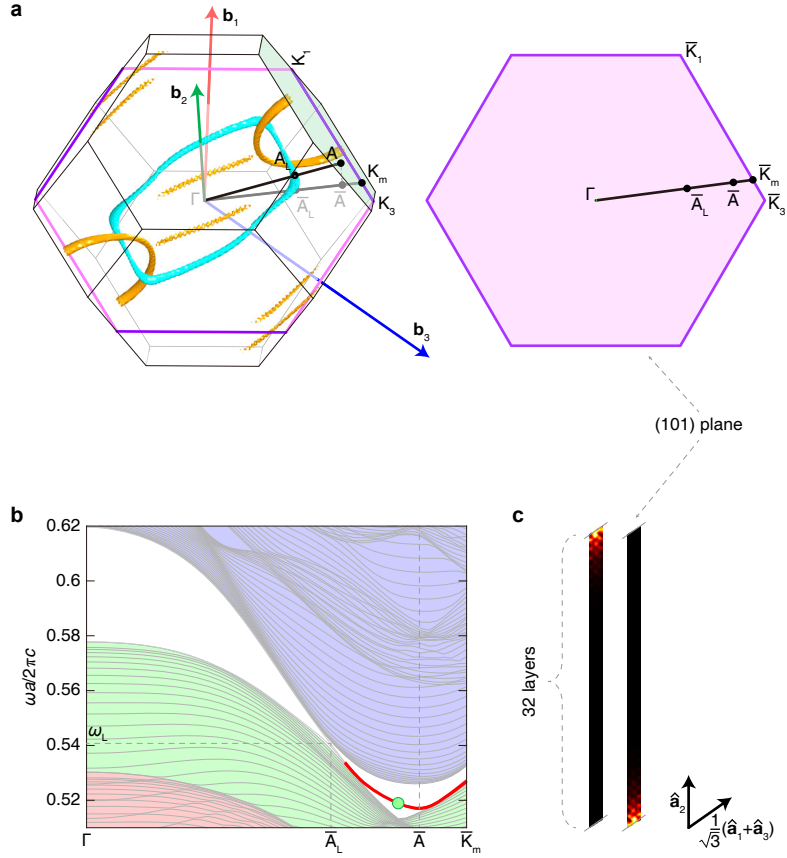

**Fig. 5. Surface states on the interface between the double diamond structure and PEC boundary. a**, (Left) Bulk Brillouin zone with the nodal link. The magenta-colored hexagon is the (101) surface. The black-colored path ΓA passes the cyan nodal line. Its projection onto the (101) plane is the gray-colored path. (Right) Surface Brillouin zone extracted from the left figure. $\bar{A}_L$ and $\bar{A}$ are the projected points of $A_L$ and A onto the (101) plane, respectively. **b**, Simulated band structure of the 32-layered double diamond structure. The overlapped two bands which correspond surface states are depicted as red. $\omega_L$ is the frequency at $\bar{A}_L$. **c**, The eigenstates $\|\mathbf{H}(\mathbf{x})\|$ at the green point in **b**.

**Methods**

All Photonic band structures and eigenstates in this study, except the data in Fig. 5b-c, were obtained by the MPB[50]. To find the degeneracies, the MPB calculations were performed for a cuboid in the momentum space with $-2\pi/a \leq k_x \leq 2\pi/a$, $-2\pi/a \leq k_y \leq 2\pi/a$, and $0 \leq k_z \leq 2\pi/a$. We used 3D grids of 101 × 101 × 51 for this cuboid. The results on this region were copied by inversion symmetry. We set a tolerance as 0.0045; if a normalized frequency difference ($\Delta\omega a/2\pi c$) of adjacent two bands at **k** is smaller than the tolerance, we considered the **k** to be degenerated by the bands. After the collection of the degeneracies, only the points **k** in the first Brillouin zone were plotted and it is the nodal link shown in throughout this study.

Fig. 5b-c were calculated by COMSOL Multiphysics®. We prepared 32 FCC unit cells along $\mathbf{a}_2$-direction. PEC was imposed on two boundaries parallel both $\mathbf{a}_1$ and $\mathbf{a}_3$. Other boundaries were set as periodic to the opposite boundaries.

**Acknowledgement**

This research was undertaken using the supercomputing facilities at Cardiff University operated by Advanced Research Computing at Cardiff (ARCCA) on behalf of the Cardiff Supercomputing Facility and the HPC Wales and Supercomputing Wales (SCW) projects. We acknowledge the support of the SCW projects and Sêr Cymru II Rising Star Fellowship (80762-CU145 (East)), which are part-funded by the European Regional Development Fund (ERDF) via the Welsh Government. This research was also supported by the National Natural Science Foundation of China (Grant No. 11874431), the National Key R&D Program of China (Grant No. 2018YFA0306800), and the Guangdong Science and Technology Innovation Youth Talent Program (Grant No. 2016TQ03X688).

**Author contributions**

S.S.O. and X.Z. conceived this topic. S.S.O. supervised all this work. H.P. proposed the double diamond structure, simulated the photonic band structures, and analyzed the non-Abelian topological charges. S.W. reviewed all mathematical derivations. H.P. and S.S.O. wrote the manuscript. All authors contributed to discussion of the data and the manuscript.

**Competing interests**

The authors declare that they have no competing interests.

**Additional information**

Correspondence and requests for materials should be addressed to S.S.O.

Supplementary information for

# "Non-Abelian charged nodal links in dielectric photonic crystal"

Haedong Park[1], Stephan Wong[1], Xiao Zhang[2], Sang Soon Oh[1,*]

[1]School of Physics and Astronomy, Cardiff University, Cardiff CF24 3AA, United Kingdom

[2]School of Physics, Sun Yat-sen University, Guangzhou 510275, China

*Email: ohs2@cardiff.ac.uk

# Contents

# Section 1. Degeneracy distributions of double diamond photonic crystal without breaking symmetries

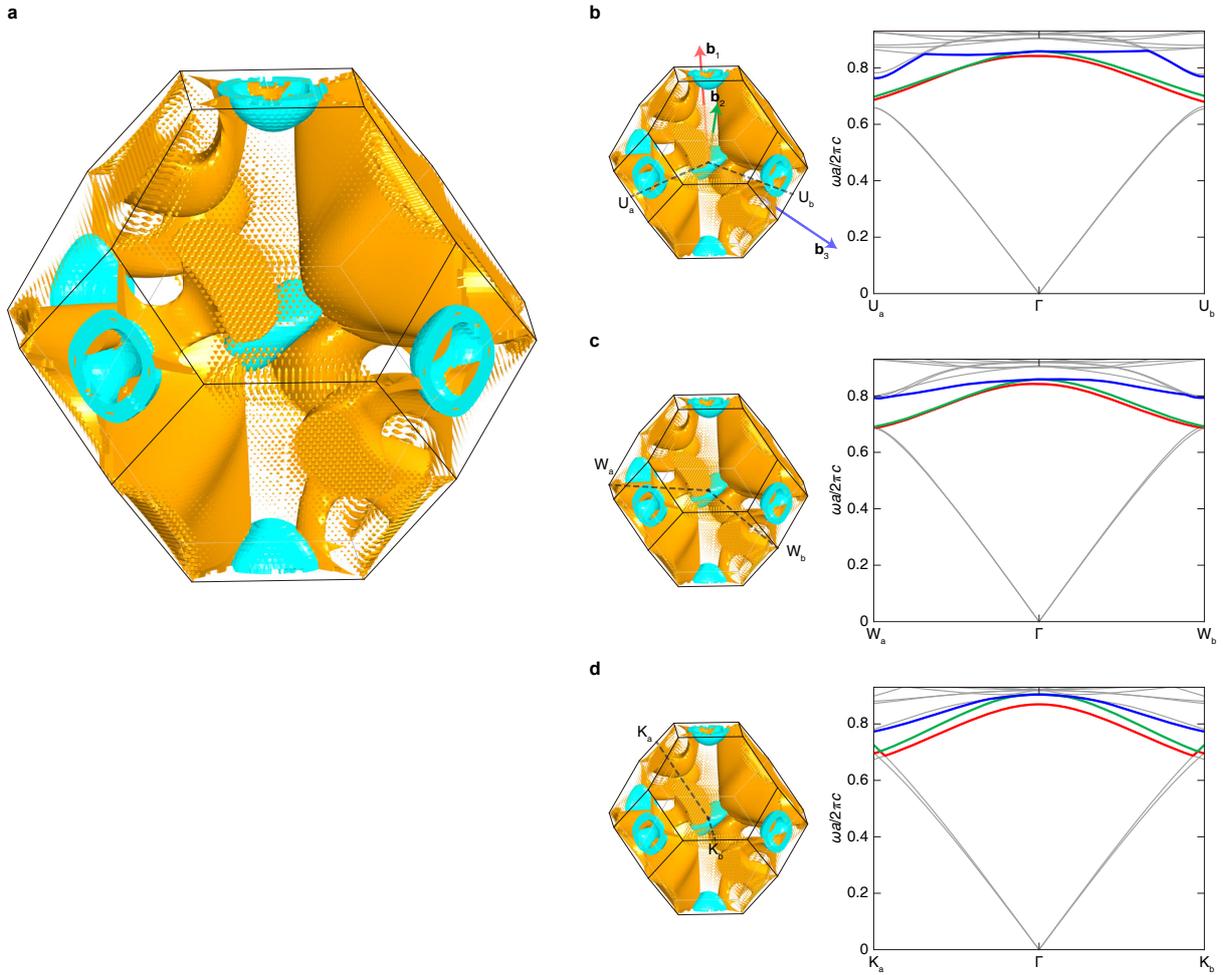

**Fig. S1. Behavior of degeneracies of double diamond photonic crystal in space group of** $Pn\bar{3}m$ **(No. 224).** Here, we used $A_1 = A_2 = A_3 = 1$, $\gamma = 0$, $f_c = 1.95$, and 16.0 for dielectric permittivity of both structures. **b**, Degeneracies of the double diamond photonic crystal in 3D momentum space. The orange and cyan nodal lines are respectively formed by the 3rd and 4th bands and 4th and 5th bands in the band structure. **c-e**, Band structures (right) along the several paths (left).

## Section 2. Symmetries of the double diamond structure

Equation (1) in the main text has four sine terms and their coefficients are $1$, $A_1$, $A_2$, and $A_3$, respectively. If $\gamma = 0$, the inner functions of the sine terms are $X_1 + X_2 + X_3$, $-X_1 + X_2 + X_3$, $X_1 - X_2 + X_3$, and $X_1 + X_2 - X_3$ in sequence. If we apply an operation of the space group $Fd\bar{3}m$ (No. 227) to the equation and sort the resulting sine terms in order of the above sequence, the sequence of the coefficients on the sine terms changes. For example, if we consider a 2-fold screw axis which consists of a 2-fold rotation axis passing through $[1/8, 0, 3/8]a$ and a translation along $[0, 1/2, 0]a$, this screw axis transforms the equation into

$$f(\mathbf{x}) = A_2 \sin(X_1 + X_2 + X_3) + A_3 \sin(-X_1 + X_2 + X_3) + \sin(X_1 - X_2 + X_3) + A_1 \sin(X_1 + X_2 - X_3) \quad \text{(S1)}$$

The coefficients of above equation show different sequence from equation (1) in the main text. If $A_2 \neq 1$ or $A_1 \neq A_3$, this level function does not satisfy the 2-fold screw axis symmetry. The level surface of each single diamond structure (Fig. 1a in the main text) becomes anisotropic along all directions by setting different coefficients on all sine terms of equation (1) in the main text. If an operation preserves the coefficients sequence, the anisotropic diamond structure would be symmetric for the operation. These operations are only inversion and translation; all the other operations change the sequence. This means that imposing the anisotropic geometry on each single diamond structure breaks all symmetries in the space group $Fd\bar{3}m$ (No. 227) except the translation and self-inversion. Here, the 'self-inversion symmetric' means that a single diamond structure itself satisfies the inversion symmetric; it is different from the 'mutual inversion symmetric' that an inversion of a single diamond structure is the other single diamond structure.

Meanwhile, if two mutual inversion symmetric and anisotropic level surfaces $f(\mathbf{x})$ and $f(-\mathbf{x})$ are given and $\gamma = 0$ still holds, this composite structure has the following translational symmetries along $\langle a/2,0,0 \rangle$ so that the primitive cubic cell with the lattice constant is $a/2$ can be used:

$$f\left(\mathbf{x} - \left[\tfrac{a}{2},0,0\right]\right) = \sin\left\{\tfrac{2\pi}{a}\left(x_1 + x_2 + x_3 - \tfrac{a}{2}\right)\right\} + A_1 \sin\left\{\tfrac{2\pi}{a}\left(-x_1 + x_2 + x_3 + \tfrac{a}{2}\right)\right\}$$

$$+ A_2 \sin\left\{\tfrac{2\pi}{a}\left(x_1 - x_2 + x_3 - \tfrac{a}{2}\right)\right\} + A_3 \sin\left\{\tfrac{2\pi}{a}\left(x_1 + x_2 - x_3 - \tfrac{a}{2}\right)\right\} = f(-\mathbf{x}) \quad (S2)$$

However, the non-zero $\gamma$ breaks the translation symmetries as follows so that the face-centered cubic (FCC) cell has to be used:

$$f\left(\mathbf{x} - \left[\tfrac{a}{2},0,0\right]\right)$$

$$= \sin\left\{\tfrac{2\pi}{a}\left(x_1 + x_2 + x_3 - \tfrac{3a\gamma}{2} - \tfrac{a}{2}\right)\right\} + A_1 \sin\left\{\tfrac{2\pi}{a}\left(-x_1 + x_2 + x_3 - \tfrac{a\gamma}{2} + \tfrac{a}{2}\right)\right\}$$

$$+ A_2 \sin\left\{\tfrac{2\pi}{a}\left(x_1 - x_2 + x_3 - \tfrac{a\gamma}{2} - \tfrac{a}{2}\right)\right\} + A_3 \sin\left\{\tfrac{2\pi}{a}\left(x_1 + x_2 - x_3 - \tfrac{a\gamma}{2} - \tfrac{a}{2}\right)\right\}$$

$$= -\sin\left\{\tfrac{2\pi}{a}\left(x_1 + x_2 + x_3 - \tfrac{3a\gamma}{2}\right)\right\} - A_1 \sin\left\{\tfrac{2\pi}{a}\left(-x_1 + x_2 + x_3 - \tfrac{a\gamma}{2}\right)\right\}$$

$$- A_2 \sin\left\{\tfrac{2\pi}{a}\left(x_1 - x_2 + x_3 - \tfrac{a\gamma}{2}\right)\right\} - A_3 \sin\left\{\tfrac{2\pi}{a}\left(x_1 + x_2 - x_3 - \tfrac{a\gamma}{2}\right)\right\}$$

$$= \sin\left\{\tfrac{2\pi}{a}\left(-x_1 - x_2 - x_3 + \tfrac{3a\gamma}{2}\right)\right\} + A_1 \sin\left\{\tfrac{2\pi}{a}\left(x_1 - x_2 - x_3 + \tfrac{a\gamma}{2}\right)\right\}$$

$$+ A_2 \sin\left\{\tfrac{2\pi}{a}\left(-x_1 + x_2 - x_3 + \tfrac{a\gamma}{2}\right)\right\} + A_3 \sin\left\{\tfrac{2\pi}{a}\left(-x_1 - x_2 + x_3 + \tfrac{a\gamma}{2}\right)\right\}$$

$$\neq f(-\mathbf{x}) \quad (S3)$$

This non-zero $\gamma$ also acts on the positions of the self-inversion symmetric points of the single diamonds. The points are $\bar{\mathbf{a}} = a/4\,\mathbf{n} \pm \gamma \mathbf{a}/2$ where the components of $\mathbf{n} = [n_1, n_2, n_3]$ are integers such that $\|\mathbf{n}\|^2$ is an odd number. The self-inversion symmetries of the single diamonds with respect to these points can be shown as follows by substituting $-\mathbf{x} + 2\bar{\mathbf{a}}$ into $\mathbf{x}$:

$$f(\pm(-\mathbf{x} + 2\bar{\mathbf{a}}))$$

$$= \sin\left[\frac{2\pi}{a}\left\{\pm(-x_1 + 2\bar{a}_1) \pm (-x_2 + 2\bar{a}_2) \pm (-x_3 + 2\bar{a}_3) - \frac{3a\gamma}{2}\right\}\right]$$

$$+ A_1 \sin\left[\frac{2\pi}{a}\left\{\mp(-x_1 + 2\bar{a}_1) \pm (-x_2 + 2\bar{a}_2) \pm (-x_3 + 2\bar{a}_3) - \frac{a\gamma}{2}\right\}\right]$$

$$+ A_2 \sin\left[\frac{2\pi}{a}\left\{\pm(-x_1 + 2\bar{a}_1) \mp (-x_2 + 2\bar{a}_2) \pm (-x_3 + 2\bar{a}_3) - \frac{a\gamma}{2}\right\}\right]$$

$$+ A_3 \sin\left[\frac{2\pi}{a}\left\{\pm(-x_1 + 2\bar{a}_1) \pm (-x_2 + 2\bar{a}_2) \mp (-x_3 + 2\bar{a}_3) - \frac{a\gamma}{2}\right\}\right]$$

$$= -\sin\left[\frac{2\pi}{a}\left\{(\pm x_1 \pm x_2 \pm x_3) + \frac{3a\gamma}{2} \mp 2(\bar{a}_1 + \bar{a}_2 + \bar{a}_3)\right\}\right]$$

$$- A_1 \sin\left[\frac{2\pi}{a}\left\{(\mp x_1 \pm x_2 \pm x_3) + \frac{a\gamma}{2} \mp 2(-\bar{a}_1 + \bar{a}_2 + \bar{a}_3)\right\}\right]$$

$$- A_2 \sin\left[\frac{2\pi}{a}\left\{(\pm x_1 \mp x_2 \pm x_3) + \frac{a\gamma}{2} \mp 2(\bar{a}_1 - \bar{a}_2 + \bar{a}_3)\right\}\right]$$

$$- A_3 \sin\left[\frac{2\pi}{a}\left\{(\pm x_1 \pm x_2 \mp x_3) + \frac{a\gamma}{2} \mp 2(\bar{a}_1 + \bar{a}_2 - \bar{a}_3)\right\}\right]$$

$$= -\sin\left[\frac{2\pi}{a}\left\{(\pm x_1 \pm x_2 \pm x_3) - \frac{3a\gamma}{2} \mp \frac{a}{2}(n_1 + n_2 + n_3)\right\}\right]$$

$$- A_1 \sin\left[\frac{2\pi}{a}\left\{(\mp x_1 \pm x_2 \pm x_3) - \frac{a\gamma}{2} \mp \frac{a}{2}(-n_1 + n_2 + n_3)\right\}\right]$$

$$- A_2 \sin\left[\frac{2\pi}{a}\left\{(\pm x_1 \mp x_2 \pm x_3) - \frac{a\gamma}{2} \mp \frac{a}{2}(n_1 - n_2 + n_3)\right\}\right]$$

$$- A_3 \sin\left[\frac{2\pi}{a}\left\{(\pm x_1 \pm x_2 \mp x_3) - \frac{a\gamma}{2} \mp \frac{a}{2}(n_1 + n_2 - n_3)\right\}\right]$$

$$= f(\pm \mathbf{x}) \quad \quad (S4)$$

Therefore, the non-zero $\gamma$ splits the self-inversion symmetric points of the two single diamonds along $[1, 1, 1]$-direction which were identical under $\gamma = 0$.

## Section 3. Topological nature on the nodal lines outside the link

In our results shown in Fig. 1b in the main text, there are another nodal lines outside the link. In the first Brillouin zone, totally six segments exist. Due to the periodicity of the first Brillouin zone, each segment extends infinitely and connects to other segments. Their connectivities are marked in Fig. S2a. An end of a segment bonds onto the end of other segment with the same number. Then, all these segments are classified as only two groups containing points [A1, A2, A3] and [B1, B2, B3], respectively, as shown in Fig. S2a.

Their topological nature can be determined by using the same method mentioned in the main text. At first, a closed loop is considered which encircles a section in the nodal line (Fig. S2a-b). Along this loop, the correlation vectors $\mathbf{C}_n$ are calculated. Gathering all tails of these vectors generates Fig. S2c. Therefore, these nodal lines are also topologically nontrivial and their topological nature can be considered as non-Abelian charge, $-i\sigma_3 \mapsto \mathbf{k}$.

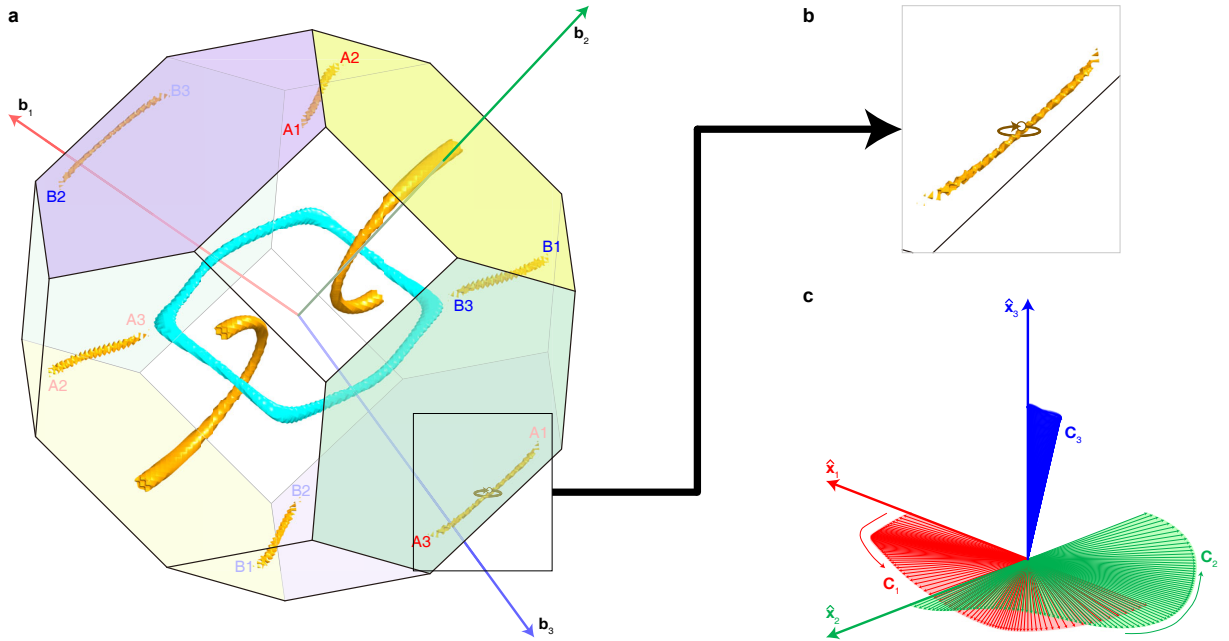

**Fig. S2. Topological analysis on the nodal lines outside the link. a**, Connectivity of the nodal lines. **b**, A closed loop enclosing the nodal line. **c**, Correlations $\mathbf{C}_n$ whose vector tails are collected at the origin to see their topological charges.

# Section 4. Nodal chain by the double diamond photonic crystal

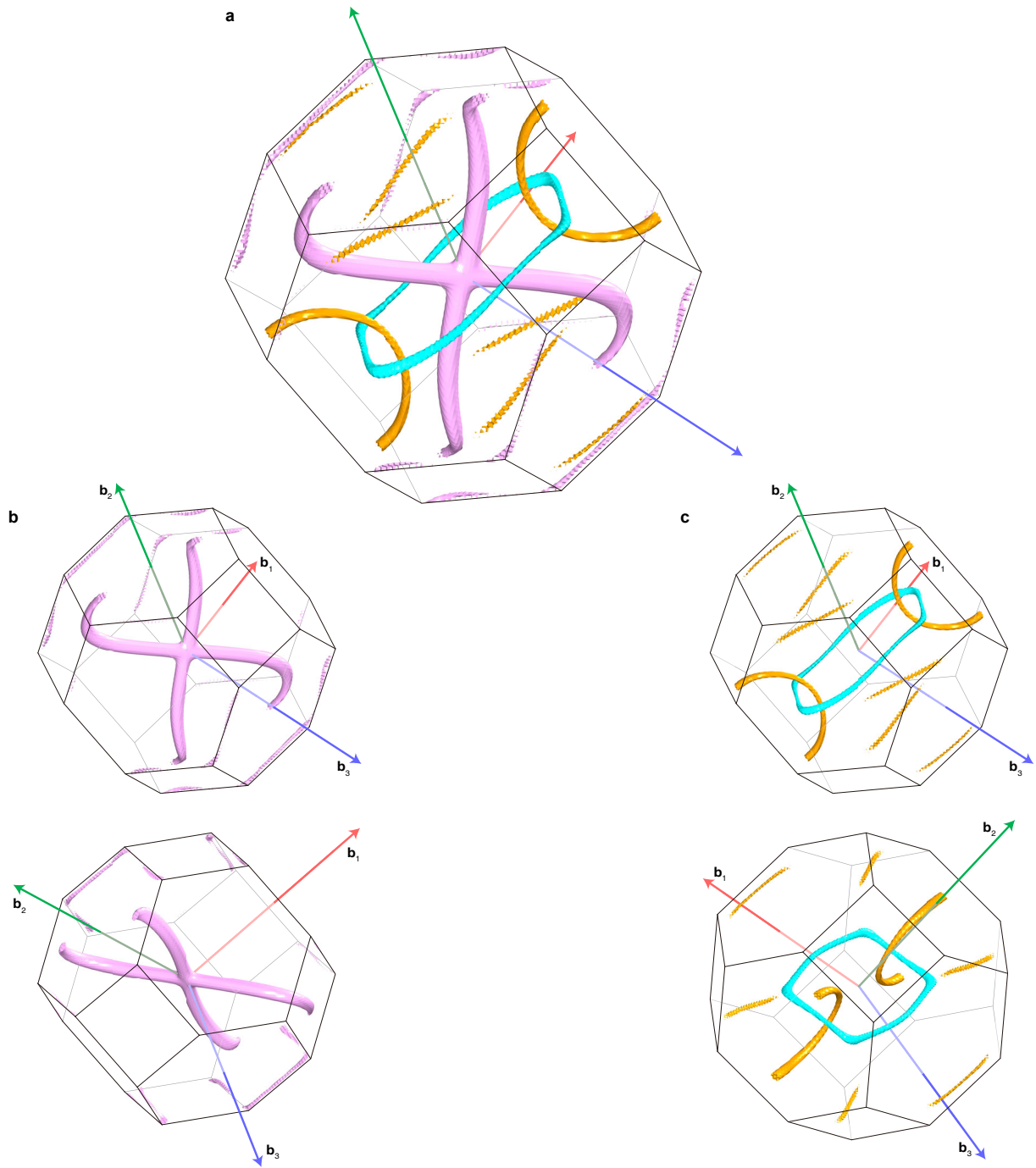

**Fig. S3. a**, Summary of the nodal link and the chain by our double diamond photonic crystal. The pink nodal chain, cyan and orange rings are formed by the 1st and 2nd bands, 3rd and 4th bands, and 4th and 5th bands in the band structure, respectively. **b**, Plot of only nodal chain. **c**, Plot of only nodal link discussed throughout the main text.

## Section 5. Sign convention of the eigenstates

The correlation $\mathbf{C}_n(\mathbf{k})$ whose components are defined by $[\mathbf{C}_n(\mathbf{k})]_m = \langle u_{\mathbf{k}_0}^m | u_{\mathbf{k}}^n \rangle$ is the main concept which explains the non-Abelian quaternion charges throughout this study. However, there is an important preparation to calculate the correlations. Basically, all eigenstates obtained from an eigenvalue problem exhibit the sign ambiguity, i.e., if a state $|u_{\mathbf{k}}^n\rangle$ is an eigenstate of a Hamiltonian at $\mathbf{k}$, $-|u_{\mathbf{k}}^n\rangle$ is also the eigenstate. Thus, if we know $|u_{\mathbf{k}}^n\rangle$ with its correct sign, the sign of $|u_{\mathbf{k}+\Delta\mathbf{k}}^n\rangle$ can be determined such that $\text{Re}\langle u_{\mathbf{k}}^n | u_{\mathbf{k}+\Delta\mathbf{k}}^n\rangle > 0$.

The problem is how to guarantee that the signs of $|u_{\mathbf{k}_0}^n\rangle$ at the starting points of the several closed loops are determined under the same convention. For example, two closed loops in Fig. 2a-d in the main text are in completely different locations. In addition, though the loop in Fig. 3a in the main text encloses both the two nodal lines to discuss $-i\sigma_2 = (-i\sigma_3)(-i\sigma_1)$, this loop does not intersect any of two loops in Fig. 2a-d in the main text. If the sign convention is not clearly arranged in this situation, an inconsistency may occur such as sign mismatching that the topological charge of the loop Fig. 3a in the main text is not $-i\sigma_2$ but $i\sigma_2$ so that the non-Abelian relations cannot be used.

To elucidate this ambiguity, we use a common reference point $\mathbf{k}_r$ (also known as the base point[1,2]) and lines which connect $\mathbf{k}_r$ and $\mathbf{k}_0$ of each loop as shown in Fig. S4. When the topological charge on the orange path in Fig. S4 is investigated, $|u_{\mathbf{k}_r}^n\rangle$ is considered above all. First, the sign of $|u_{\mathbf{k}_r+\Delta\mathbf{k}}^n\rangle$ is determined such that satisfies the criterion $\text{Re}\langle u_{\mathbf{k}_r}^n | u_{\mathbf{k}_r+\Delta\mathbf{k}}^n\rangle > 0$ mentioned in the first paragraph in this section. For all the remaining points on the line, the sign of $|u_{\mathbf{k}+\Delta\mathbf{k}}^n\rangle$ is determined by $\text{Re}\langle u_{\mathbf{k}}^n | u_{\mathbf{k}+\Delta\mathbf{k}}^n\rangle > 0$. Finally, the sign of $|u_{\mathbf{k}_0}^n\rangle$ is determined. Same convention is applied to the cyan path in Fig. S4.

The difference of the reference point method, compared to Refs. 1,2, is that this study does not need the reverse line from $\mathbf{k}_0$ to $\mathbf{k}_r$ because the correlation is defined with respect to not $|u_{\mathbf{k}_r}^n\rangle$ but $|u_{\mathbf{k}_0}^n\rangle$ and the aim of the reference point and the line in this study is only determining the signs of $|u_{\mathbf{k}_0}^n\rangle$ with the same convention.

The correlation $C_n(\mathbf{k})$ is calculated from the sign-corrected $|u_{\mathbf{k}}^n\rangle$ according to this sign convention. All discussions on the non-Abelian topological charges in this study were done based on this preparation. All the preparations use a common $\mathbf{k}_r$. All closed loops shown in figures in this study omit the lines between $\mathbf{k}_r$ and $\mathbf{k}_0$ and all plots on the correlations show $C_n(\mathbf{k})$ for only after $\mathbf{k} = \mathbf{k}_0$.

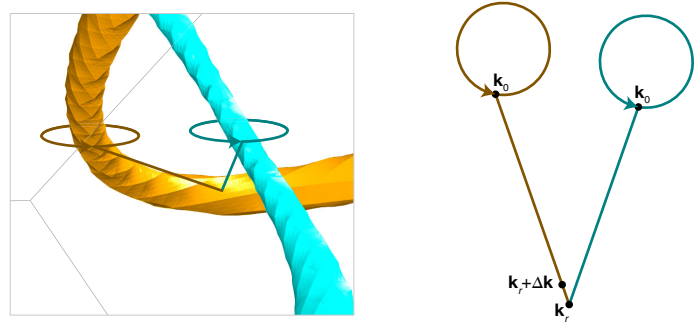

**Fig. S4. Reference point for the sign convention.** The common reference point is denoted as $\mathbf{k}_r$. The lines which connect $\mathbf{k}_r$ and $\mathbf{k}_0$ are also shown. All closed loops shown throughout this study are the results by omitting the reference point and the lines.

## Section 6. Discussions on signs of the topological charge j

From the closed loop shown in Fig. 3a in the main text, we mentioned its topological charge is $-i\sigma_2 = \mathbf{j}$. This is consistent with the relation $-i\sigma_2 = (-i\sigma_3)(-i\sigma_1)$ due to the first and second half of the loop circle around cyan $(-i\sigma_1)$ and orange $(-i\sigma_3)$ nodes, respectively. Then, we questioned about whether the topological charge is flipped from $\mathbf{j}$ to $-\mathbf{j}$ by changing this scanning sequence. If so, the topological charge $\mathbf{j}$ strongly reflects the non-Abelian nature of the topological charges.

The first variation on the loop is scanning around the orange node followed by the cyan node without flipping the winding direction (Fig. S5a). Keeping the winding direction means the signs of $(-i\sigma_3)$ and $(-i\sigma_1)$ remain. The scanning sequence makes a composite relation $(-i\sigma_1)(-i\sigma_3)$ so we can expect its topological charge is $i\sigma_2 = -\mathbf{j}$. Indeed, the simulated correlations $\mathbf{C}_3$ and $\mathbf{C}_1$ reveal the rotations along the opposite direction which corresponds to $i\sigma_2 = -\mathbf{j}$ (Fig. S5b).

The second variation on the loop is same to the original loop in Fig. 3a in the main text except the opposite winding direction (Fig. S5c). Due to this winding, the topological charges around the orange and cyan nodes are $i\sigma_3$ and $i\sigma_1$, respectively. Their composition is expressed as $(i\sigma_1)(i\sigma_3)$, thus $i\sigma_2 = -\mathbf{j}$. The correlations in Fig. S5d also match to this expectation.

The final variation on the loop envelops the cyan node first and its winding direction is also opposite to the original loop (Fig. S5e). The expectation $(i\sigma_3)(i\sigma_1)$ and the correlations (Fig. S5f) commonly arrive at $-i\sigma_2 = \mathbf{j}$.

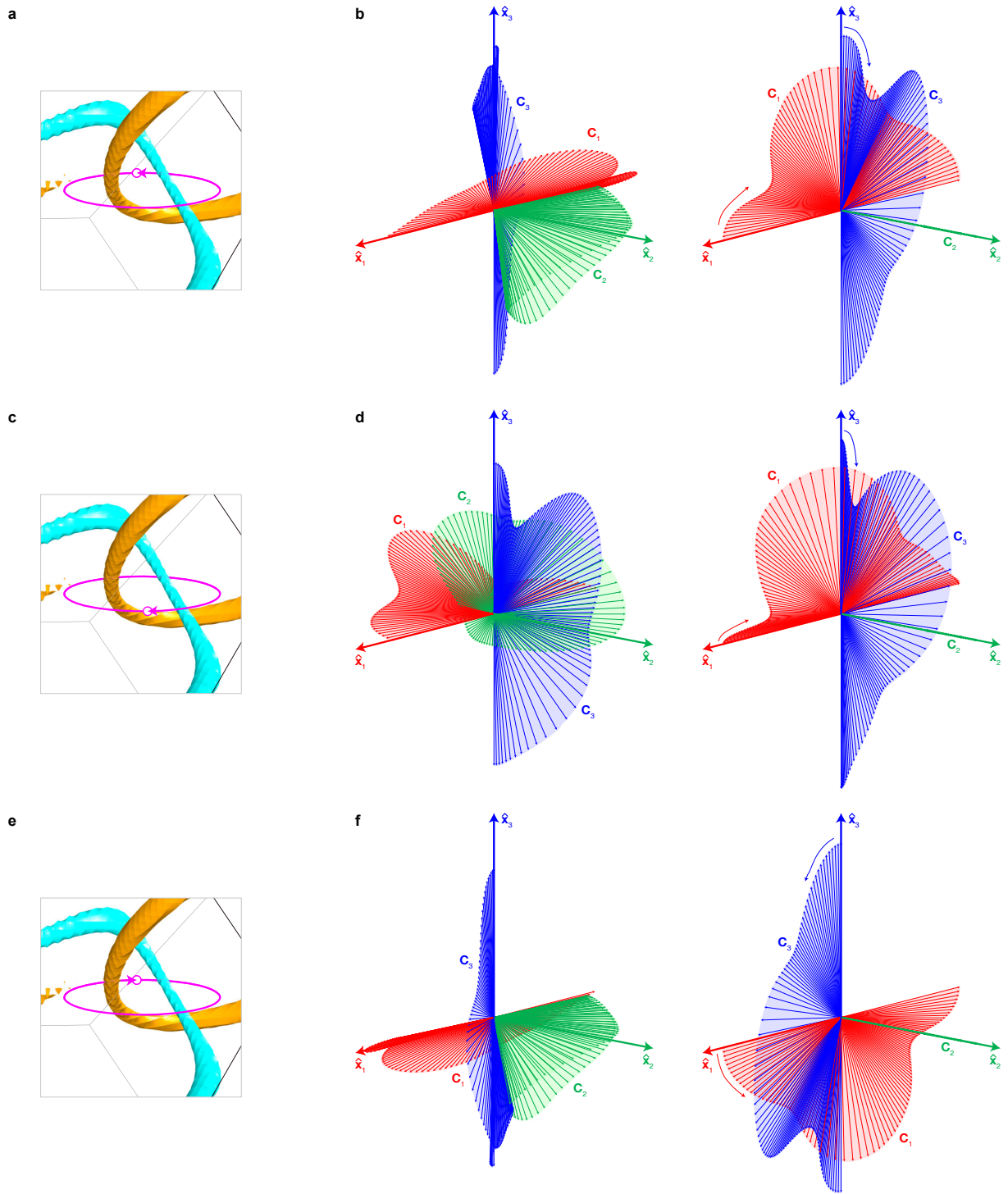

**Fig. S5. Sign of the topological charge j by the variations on the closed loop. a**, **c**, and **e**, Three variations on the closed loop in Fig. 3a in the main text. **b**, **d**, and **f**, The correlations $C_n$. In each panel, the right figure is the calibrated result of the left figure.

## Section 7. Determination of the topological charge j without calibration

In the main text, the topological charge $-i\sigma_2$ was derived by the calibration using the rotation matrix which decreases the angle between $\mathbf{C}_2$ and $\hat{\mathbf{x}}_2$-axis. Here, this charge is derived without the calibration. Because the closed loop in Fig. 3a in the main text contains both cyan and orange rings, a rotation matrix can be written as a composition form:

$$R_{31}(\alpha) = R_{12}\big(g_{12}(\alpha)\big) R_{23}\big(g_{23}(\alpha)\big) = e^{(g_{12}/2)L_3} e^{(g_{23}/2)L_1} \tag{S5}$$

Unlike the case of $(-i\sigma_3)$ and $(-i\sigma_1)$, the input variables in $R_{12}$ and $R_{23}$ are $g_{12}(\alpha)$ and $g_{23}(\alpha)$, respectively, rather than $\alpha$ where these two are arbitrary continuous functions whose values are $0$ and $2\pi$ for $\alpha = 0$ and $2\pi$, respectively. Introducing these two functions is justified because the circular closed loop is not centered at a point on the rings; both ring points which meet the closed loop surface are slightly lopsided near to the loop.

When the ring points are positioned as shown in Fig. S6a, $R_{31}(\alpha)$ for $\alpha \in [0, 2\pi]$ transforms each unit vector $\hat{\mathbf{x}}_n$. Their plots in Fig. S6b show similar features to the Fig. 3c in the main text. First, the transformations on $\hat{\mathbf{x}}_1$ and $\hat{\mathbf{x}}_3$ exhibit $\pi$-disclinations while $\hat{\mathbf{x}}_2$ does not, i.e., $R_{31}(0)\hat{\mathbf{x}}_1 = -R_{31}(2\pi)\hat{\mathbf{x}}_1$, $R_{31}(0)\hat{\mathbf{x}}_3 = -R_{31}(2\pi)\hat{\mathbf{x}}_3$, $R_{31}(0)\hat{\mathbf{x}}_2 = +R_{31}(2\pi)\hat{\mathbf{x}}_2$. Second, $R_{31}(\alpha)\hat{\mathbf{x}}_2$ do not stay near around the original $\hat{\mathbf{x}}_2$-axis. This large angle between $R_{31}(\alpha)\hat{\mathbf{x}}_2$ and $\hat{\mathbf{x}}_2$-axis was also the reason that we used the calibration in the main text. Finally, $R_{31}(\alpha)\hat{\mathbf{x}}_3$ are not on the $\hat{\mathbf{x}}_1\hat{\mathbf{x}}_3$ plane. In the main text, this was adjusted by the calibration.

If the two ring points are merged at the center of the circle (Fig. S6c), we can use $g_{12}(\alpha) = g_{23}(\alpha) = \alpha$. The result still shows the above features; the $\pi$-disclinations and the large angles between $R_{31}(\alpha)\hat{\mathbf{x}}_2$ and $\hat{\mathbf{x}}_2$-axis. And the distributions of $R_{31}(\alpha)\hat{\mathbf{x}}_3$ which are not on the $\hat{\mathbf{x}}_1\hat{\mathbf{x}}_3$ plane are also observed, though the center of the distributions is moved onto the plane.

Likewise, Fig. S6b-d are induced from equation (S5) and their tendencies are similar to Fig. 3c in the main text. Fortunately, the lift of equation (S5) in the double cover Spin(3)

$$\bar{R}_{31}(\alpha) = e^{-i(g_{12}/4)\sigma_3} e^{-i(g_{23}/4)\sigma_1} \tag{S6}$$

gives the consistent topological charge as $\bar{R}_{31}(2\pi) = (-i\sigma_3)(-i\sigma_1) = -i\sigma_2$.

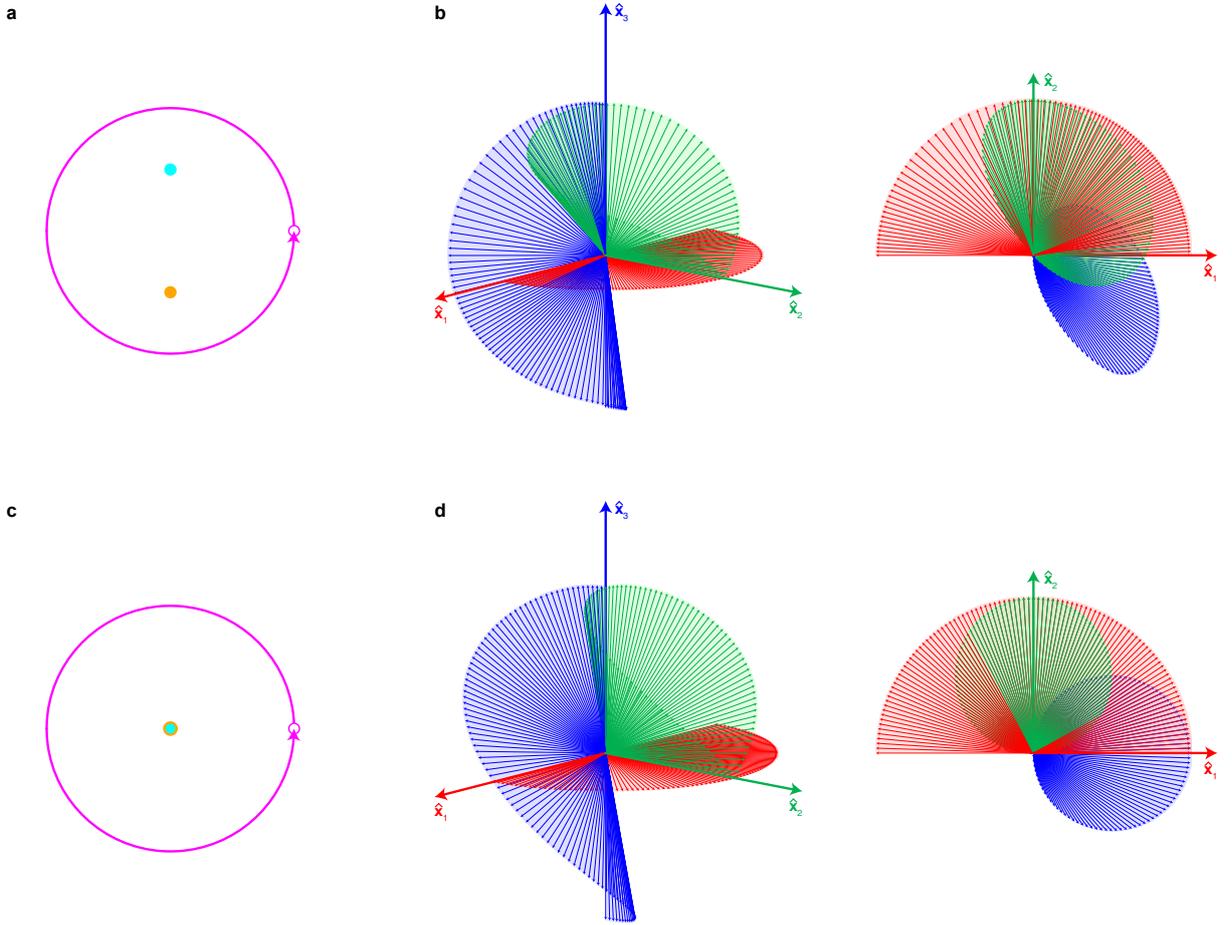

**Fig. S6. Topological charge j by the rotation matrix $R_{31}(\alpha)$. a**, A closed loop which encircles both two nodal lines. The distances between both nodes and the loop center are half of the radius of the loop. **b**, Plots of $R_{31}(\alpha)\hat{\mathbf{x}}_n$. The red, green, and blue colors correspond to $n = 1, 2,$ and $3$. **c**, A closed loop with merged two nodal points. **d**, Plots of $R_{31}(\alpha)\hat{\mathbf{x}}_n$ on the case of **c**.

# Section 8. Determination of the topology of the nodal lines using Berry-Wilczek-Zee connection

We start this discussion with the generalized Wilson operator.[1,3] If we consider a closed loop enclosing a nodal line and it is parametrized by $\alpha \in (0, 2\pi]$, the generalized Wilson operator is written as[1,3]

$$W = \exp \theta_{pq} = \exp\left\{\oint_{\Gamma(\alpha)} [\mathbf{A}(\mathbf{k})]_{pq} \cdot d\mathbf{k}\right\} = \exp\left\{\oint_{\Gamma(\alpha)} [A(k)]_{pq} dk\right\} \quad (S7)$$

where

$$[A(k)]_{pq} = \left\langle u_{\mathbf{k}}^p \left| \frac{\partial}{\partial k} \right| u_{\mathbf{k}}^q \right\rangle = \int_{cell} (u_{\mathbf{k}}^p)^* \cdot \frac{\partial u_{\mathbf{k}}^q}{\partial k} d^3\mathbf{x} \quad (S8)$$

is the Berry-Wilczek-Zee (BWZ) connection.[4] The sub- and superscripts $p$ and $q$ are the band numbers ($p, q = 3, 4, 5$ for 3rd, 4th, and 5th bands, respectively) and $u_{\mathbf{k}}^p$ is a periodic part of the magnetic field eigenstate $\mathbf{H}^p(\mathbf{x}) = u_{\mathbf{k}}^p e^{i\mathbf{k}\cdot\mathbf{x}}$ of the $p$th band such that $\langle u_{\mathbf{k}}^p | u_{\mathbf{k}}^q \rangle = \delta_{pq}$.

From the results shown in Fig. 1b in the main text, we calculated $\theta_{34}(\alpha)$ and $\theta_{45}(\alpha)$ as plotted in Fig. S7a-b, respectively. From these results, the topological charges can be assigned on each nodal line; for the closed loop encircling the nodal line by the 3rd and 4th (4th and 5th) bands, only $\theta_{34}(\alpha)$ ($\theta_{45}(\alpha)$) varies by $-\pi$ while the other remains zero.

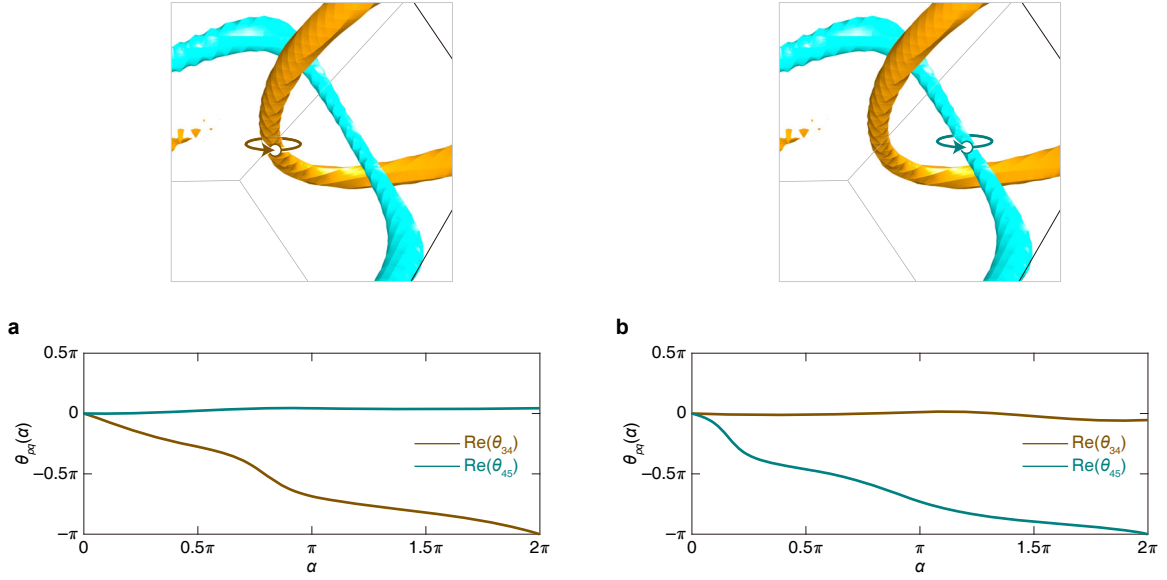

**Fig. S7. Topological nature of the nodal lines using Berry-Wilczek-Zee connection. a**, and **b**, The $\theta_{34}(\alpha)$ and $\theta_{45}(\alpha)$ with respect to $\alpha \in (0, 2\pi]$ for the nodal line by the 3rd and 4th (**a**) and 4th and 5th bands (**b**). Only real part of each value is plotted because their imaginary parts are negligibly small.